\documentclass[
 aip,
 amsmath,amssymb,
 reprint,%
]{revtex4-1}

\usepackage{graphicx}
\usepackage{dcolumn}
\usepackage{bm}
\usepackage[normalem]{ulem}%

\usepackage[utf8]{inputenc}
\usepackage[T1]{fontenc}
\usepackage{mathptmx}
\usepackage{etoolbox}

\makeatletter
\def\@email#1#2{%
 \endgroup
 \patchcmd{\titleblock@produce}
  {\frontmatter@RRAPformat}
  {\frontmatter@RRAPformat{\produce@RRAP{*#1\href{mailto:#2}{#2}}}\frontmatter@RRAPformat}
  {}{}
}%
\makeatother

\usepackage{bm}
\usepackage{float}
\usepackage{xcolor}

\newcommand{\s}{$S_{21}$}
\newcommand{\qc}{Q$_C$}

\begin{document}

\preprint{AIP/123-QED}

\title{Tantalum Damascene Coplanar Waveguide Resonators Fabricated Using 300~mm Scale Processes}

\author{Ekta Bhatia}
\email{ebhatia@ny-creates.org}
\affiliation{NY Creates, Albany, New York 12203, United States}%

\author{Yingge Du}
\email{yingge.du@pnnl.gov}
\affiliation{Physical and Computational Sciences Directorate Pacific Northwest National Laboratory, Richland, Washington 99354, United States}%

\author{Krishna P Koirala}
\affiliation{Physical and Computational Sciences Directorate Pacific Northwest National Laboratory, Richland, Washington 99354, United States}
\affiliation{Materials Department, University of California, Santa Barbara, California 93106, United States}%

\author{Chung Kow}
\affiliation{NY Creates, Albany, New York 12203, United States}%

\author{Mingzhao Liu}
\email{mzliu@bnl.gov}
\affiliation{Center for Functional Nanomaterials, Brookhaven National Laboratory, Upton, New York 11973, United States}%

\author{Juan Macy}
\affiliation{National Security Directorate, Pacific Northwest National Laboratory, Richland, Washington 99354, United States}%

\author{Tharanga R. Nanayakkara}
\affiliation{Center for Functional Nanomaterials, Brookhaven National Laboratory, Upton, New York 11973, United States}%

\author{Francisco Ponce}
\email{francisco.ponce@pnnl.gov}
\affiliation{National Security Directorate, Pacific Northwest National Laboratory, Richland, Washington 99354, United States}

\author{Satyavolu S. Papa Rao}
\email{spaparao@ny-creates.org}
\affiliation{NY Creates, Albany, New York 12203, United States}%

\author{Drew J Rebar}%
\affiliation{National Security Directorate, Pacific Northwest National Laboratory, Richland, Washington 99354, United States}%

\author{Peter V. Sushko}
\affiliation{Physical and Computational Sciences Directorate Pacific Northwest National Laboratory, Richland, Washington 99354, United States}%

\author{Brent A VanDevender}
\affiliation{National Security Directorate, Pacific Northwest National Laboratory, Richland, Washington 99354, United States}%

\author{Chongmin Wang}
\affiliation{Environmental Molecular Sciences Laboratory, Pacific Northwest National Laboratory, Richland, Washington 99354, United States}%

\author{Marvin G Warner}
\affiliation{National Security Directorate, Pacific Northwest National Laboratory, Richland, Washington 99354, United States}%

\author{Zhihao Xiao}
\affiliation{NY Creates, Albany, New York 12203, United States}%

\date{\today}

\begin{abstract}
Surface oxides contribute to losses in superconducting transmon devices resulting in degraded performance. We explore the use of the damascene process to replace the sidewall native oxide of a device with a metal/substrate interface. We simulate sidewall oxidation by burying an oxide layer during fabrication. We observe a modest improvement between the two types of devices, which is suggestive of a reduction in the surface participation ratio.
\end{abstract}

\maketitle

\section{Introduction}
Superconducting coplanar waveguide (CPW) resonators are a critical component in the transmon qubit architecture and a valuable tool in understanding associated loss mechanisms. Improvements in the design and fabrication of such CPW resonators, as well as in their cryogenic characterization methodologies, have speeded advances in qubit fabrication, and in qubit coherence times (T$_1$), due to the relative simplicity of resonators vis-a-vis qubits. Studies on device design changes \cite{Muller2019, Bu2025}, constituent materials\cite{Place2021, Premkumar2021,  Verjauw2021, Murthy2022, Dhas2025}, and fabrication processes \cite{Place2021, Lozano2024, Wang2022}, including surface \cite{Bal2024, Verjauw2021, Mergenthaler2021, Chang2025, Zhou2024}, band-gap \cite{McEwen2024} and phonon \cite{Chen2024} engineering have yielded demonstrable benefits to qubit performance. Additionally, studies focusing on the substrate indicated that the performance of the devices improved with trenching \cite{Calusine2018}, degraded with substrate impurities \cite{Zhang2024}, and is inconsistent with substrate specific RF losses \cite{Turiansky2024, Read2023}.

An in-depth study into TLS losses in resonators observed that both the bulk and surface TLS at the various device interfaces greatly affect losses in Ta devices\cite{Crowley2023}. This is attributed to the fraction of the electric field energy residing in those interfaces \cite{Wang2015, McRae2020}, the surface participation ratio (SPR), which has been shown to directly affect qubit performance in 3D cavities\cite{Wang2015}. A study on surface oxides shows that Ta forms a more stable and less complex native oxide, Ta$_2$O$_5$, compared to Nb\cite{McLellan2023}. This suggests improving device performance by further modification of the surface to remove the oxide entirely by capping the surface layer \cite{Zhou2024, Chang2025}. However, this does not address the sidewall oxidation of the devices. 

This work focuses on damascene CPW resonators created by trench formation, metal deposition, and chemical mechanical planarization (CMP). IBM developed damascene copper interconnects in the 1990’s using Cu to fill tantalum-lined vias and trenches followed by CMP\cite{Andricacos1998}.  The Ta-specific damascene CMP process applied in this work for the fabrication of Ta coplanar waveguide (CPW) resonators is developed at NY Creates \cite{Bhatia2023}. This approach addresses the challenge associated with inevitable presence of native oxide on the sidewalls of etched superconductors. Using CMP, the superconductor is embedded inside the pristine silicon substrate. In this paper, we present the performance of the first Ta-based resonators fabricated using the damascene process. 

\section{Experimental}
\subsection{Damascene Resonator Fabrication}

\begin{figure}[H]
\includegraphics[width=\linewidth]{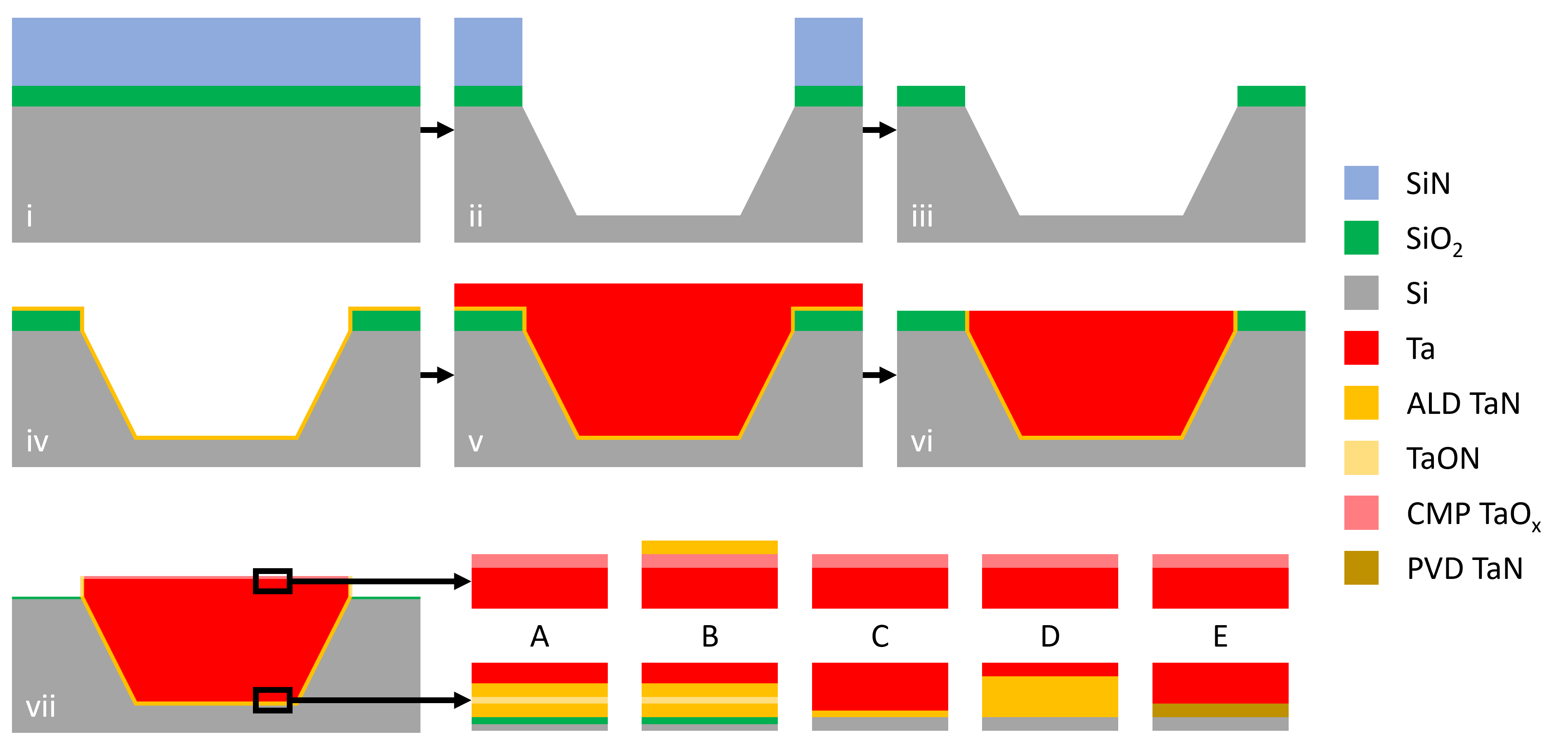}
\caption{\label{fig:Interfaces} {Damascene process flow for fabricating tantalum-based superconducting resonators.
(i–iii) Pattern definition and anisotropic Si etching using a SiO$_2$/SiN hard mask.
(iv) Deposition of a 3 nm ALD TaN seed layer.
(v) Ta sputter deposition to fill the etched Si features.
(vi) CMP removal of Ta from field regions, leaving $\alpha$-Ta
in the trenches.
(vii) Post-CMP oxide removal exposing the final recessed Ta structures.
A-E show the five device variants studied, distinguished by buried-oxide vs. pristine interfaces and different surface-treatment/capping schemes. (Dimensions not to scale)}}
\end{figure}

Tantalum-based superconducting resonators were fabricated using the damascene approach \cite{Bhatia2023}. We start with a 300~mm diameter Si wafer oriented in the (100) plane with a resistivity of 10 k$\Omega$-cm, growing a 15~nm silicon dioxide thin film and a 50~nm silicon nitride thin film, Fig. \ref{fig:Interfaces} (i). The SiO$_2$ and SiN layers form a hard mask used to define the resonator features using immersion 193~nm optical lithography and reactive ion etching (RIE). The pattern is then transferred into the Si substrate using an aqueous hydroxide-based etch to form the trenches, as shown in Fig. \ref{fig:Interfaces} (ii). The anisotropic nature of the etch exposes the \{111\} planes in Si, resulting in sloped sidewalls, which help avoid seam formation during metal deposition by sputtering. The SiN film is removed with hot phosphoric acid, while retaining the 15~nm thick SiO$_2$ layer, Fig. \ref{fig:Interfaces} (iii). From here we diverge into two types of substrate interfaces (Fig. \ref{fig:Interfaces} (iv) and Fig. \ref{fig:Interfaces} (v):
\begin{enumerate}
  \item Buried Oxide: For the oxidized-interface case, we intentionally break vacuum after deposition of the first 3~nm ALD TaN layer (Fig. \ref{fig:Interfaces} (iv)). This leads to formation of a native oxynitride on the exposed TaN surface. We hypothesized that this interfacial layer could introduce TLS loss. A second 3~nm thick ALD TaN film is then deposited to encapsulate this interfacial layer and ensure the formation of $\alpha$-Ta during the subsequent Ta sputter deposition (Fig. \ref{fig:Interfaces} (v)), without breaking vacuum again\cite{Bhatia2023}.
  \item Pristine: For the pristine-interface case, vacuum is maintained after the TaN seed-layer deposition and Ta is sputtered in situ without air exposure (Fig. \ref{fig:Interfaces} (v)). Fig. \ref{fig:Interfaces} (iv) step is not applicable to this case.  
\end{enumerate}

\begin{figure}[H]
\includegraphics[width=\linewidth]{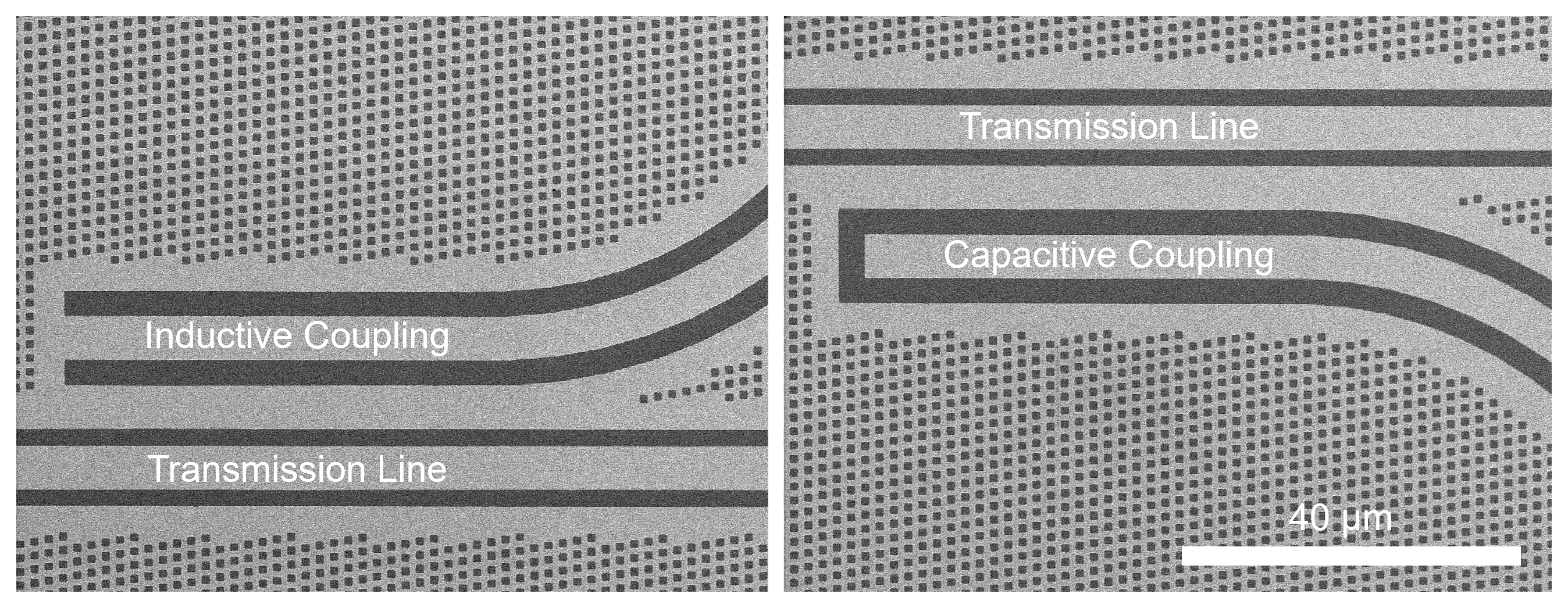}
\caption{\label{fig:Ends} Scanning electron microscope (SEM) images depicting an example/structure of an inductively (left) and capacitively (right) coupled resonator. The holes in the ground plane are a requirement for the CMP process to maintain 50\% metal coverage.}
\end{figure}

The TaN/Ta stack fills the etched trenches, resulting in a surface topography that mimics the underlying Si recesses, as shown in Fig. \ref{fig:Interfaces} (v). Finally, the Ta surface is planarized by CMP, leaving $\alpha$-Ta in the recessed regions while removing it from the SiO$_2$ field region, as shown in Fig. \ref{fig:Interfaces} (vi). The CMP process is optimized to achieve uniform planarization by appropriate choice of the slurry and pad characteristics, as well as the initial deposition thickness required to accommodate topography [26]. Note that, to achieve planarization with uniform thickness across the multiple pattern features present in a given design, it is necessary to include dummy patterns in areas that are otherwise pattern-free, as described in \cite{Bhatia2023}. Following CMP, the SiO$_2$ layer is etched away using wet etch to expose the underlying Si substrate, Fig. \ref{fig:Interfaces} (vii). Upon exposure to ambient atmosphere, the exposed Si regrows its native oxide. A total of five types of devices are fabricated with different surface treatments, Si trench depths, or TaN seed layer thicknesses:

\begin{enumerate}
\item {\bf Device A:} 
Buried oxidized interface with no additional capping layer and a a $\sim$60 nm deep Si trench.
\item {\bf Device B:} 
Buried oxidized interface with an additional 2~nm ALD TaN capping layer on top of the Ta, which seals the CMP TaO$_x$ and a $\sim$60 nm deep Si trench.
\item {\bf Device C:} 
Pristine interface with a 1~nm ALD TaN seed layer and a $\sim$48 nm deep Si trench.
\item {\bf Device D:} 
Pristine interface with a 6~nm ALD TaN seed layer and a $\sim$60 nm deep Si trench.
\item {\bf Device E:} 
Pristine interface with a 2~nm PVD TaN and a $\sim$48 nm deep Si trench.
\end{enumerate}
The successful formation of $\alpha$-Ta is confirmed by room-temperature X-ray diffraction (XRD) and resistivity measurements \cite{Bhatia2023}. The resonator chip consists of four resonators: two inductively coupled and two capacitive coupled, Fig. \ref{fig:Ends} (left, center). 

\begin{figure}[H]
\includegraphics[width=\linewidth]{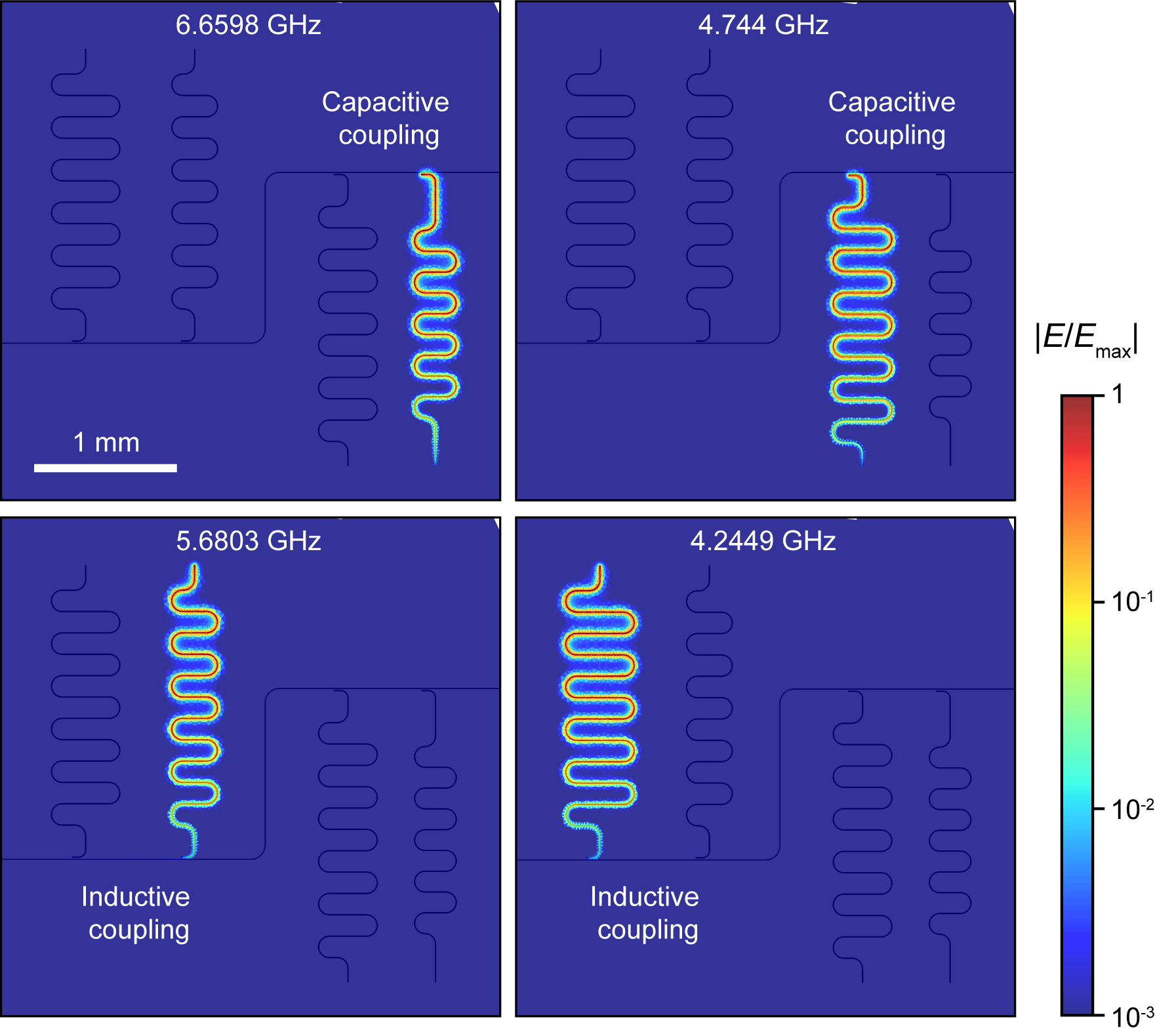}
\caption{\label{fig:E-field-simulation} {Electric field distribution profiles of the quarter-wave resonance modes of the 4 CPW resonator along the common feed line. The actual device pattern is placed to the background to help visualization.} }
\end{figure}

\subsection{Resonator Simulations}
The eigen-frequencies and coupling quality factors of the coplanar waveguide (CPW) resonators are numerically simulated using the COMSOL MultiPhysics\textregistered~RF Module. The silicon substrate is modeled as a slab with dielectric constant $\epsilon$ = 11.7  and dielectric loss tangent tan~$\delta$ = 0. The Ta superconducting CPW resonators and feed-line are modeled as a patterned plane (zero thickness) of perfect electric conductor covering the top surface of the substrate slab. Except for the center pin of the feed-line, the entire resonator device plane is grounded. The complete assembly is then enclosed in a rectangular, fully grounded box that models the shield surrounding the actual resonator device. To reduce computational complexity, the resonator pattern is simplified from the actual device by removing the “cheesing” pattern, which is introduced in the actual device to accommodate the CMP process. At either end of the feed-line, a cable-type lumped port is defined with a characteristic impedance of 50~$\Omega$. One port is set for microwave excitation and the other port is set for receiving to evaluate the dispersion of forward transmission \s. 

The simulation consists of four hanger-type resonators along the feed-line with two capacitively coupled (short to ground is located at the far-side away from the feed-line) and two inductively coupled (short to ground is located at the near the feed-line). Resonance frequencies are determined using the eigen-frequency solver. The quarter-wave resonance mode can be readily identified by its mode profiles that is geometrically confined to the CPW resonator and the eigen-frequency of a pure real number. However, the parasitic modes that are not intended by design are generally lossy and carry an imaginary part in their resonance frequency. After the eigen-frequencies are determined, the \s dispersion is evaluated for each resonator in the frequency domain, in a frequency range a few kHz around the eigen-frequency $\omega_0$. Although the CPW resonator is modeled as lossless, the \s dispersion does have a finite line-width that arises from the radiative coupling between the resonator and the feed-line. For each resonance mode, the line-width is determined by increasing the frequency resolution until the \s peak shape no longer changes with higher resolutions. Taking the peak FWHM as its line-width $\delta\omega$, the coupling quality factor can be computed as $Q_c=\frac{\omega_0}{\delta\omega}$.

Our numerical simulation gives the eigen-frequencies of the four quarter-wave CPW resonators along the common feed-line are 4.2449, 4.7440, 5.6803, and 6.6598 GHz, when kinetic inductance (KI) contribution is neglected. We confirm that for the two capacitively coupled resonators, 4.7440 and 6.6598~GHz, the mode profile calculations show the E-field node away from the feed-line, Fig. \ref{fig:E-field-simulation}. For the two inductively coupled resonators, 4.2449 and 5.6803~GHz, the mode profile calculations show the E-field nodes near the feed-line. 

The coupling quality factor between each resonator and the feed-line, \qc, is determined from the linewidth of the simulated forward transmission \s. The two capacitively coupled resonators have \qc= 3.69 × 10$^6$ ($\omega_0$ = 6.6598 GHz) and 5.68 × 10$^6$ ($\omega_0$ = 4.7440 GHz). The two inductively coupled resonators have much lower \qc, at 3.94 × 10$^5$ ($\omega_0$ = 5.6803 GHz) and 7.22 × 10$^5$ ($\omega_0$ = 4.2449 GHz). 

\subsection{Material Interface Characterization}
\begin{figure}[H]
\includegraphics[width=\linewidth]{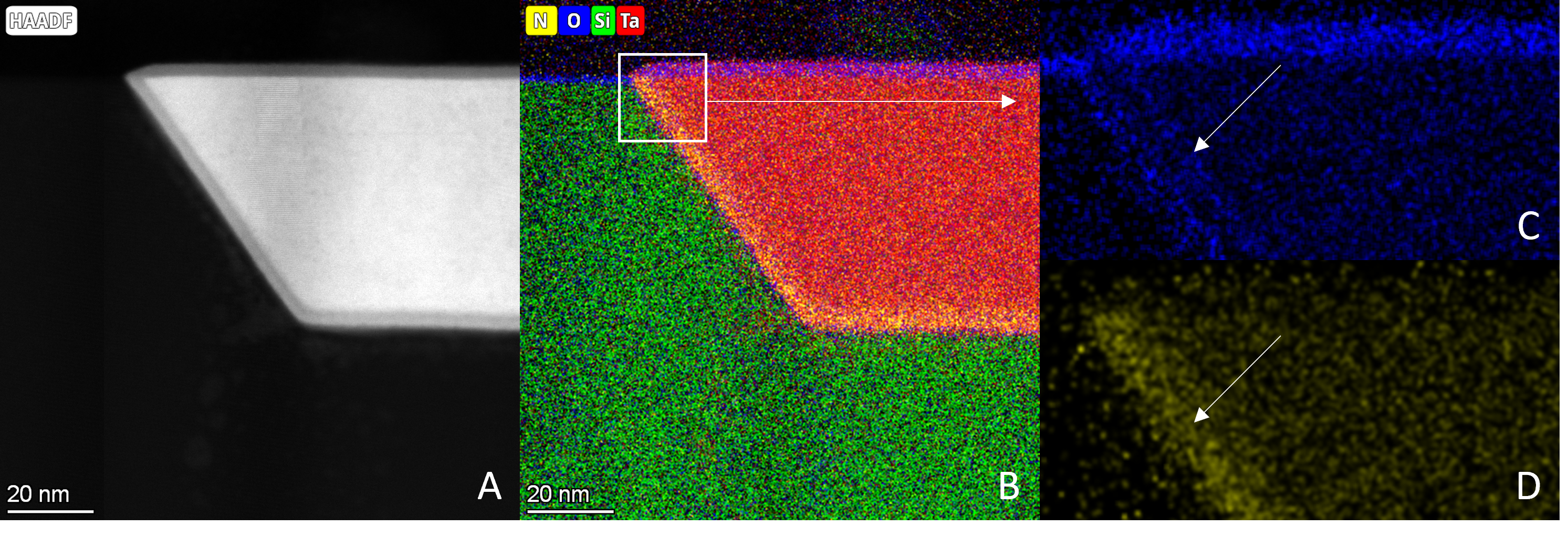}
\caption{\label{fig:Rep_A} {(A) STEM image of a Device A type showing clearly defined features relative to the Si substrate. (B) Composite elemental analysis of the device showing the N, O, Si, and Ta. Due to the overlap of the Si-K and Ta-M lines the Ta region is tinted from limitation on  elemental identification. Enhanced image of the boxed area showing (C) the trapped oxygen and (D) N layer at the Si interface. The $\sim$3 nm of TaN exposed to atmosphere (O from TaON) is clearly visible.}}
\end{figure}


\begin{figure}[H]
\includegraphics[width=\linewidth]{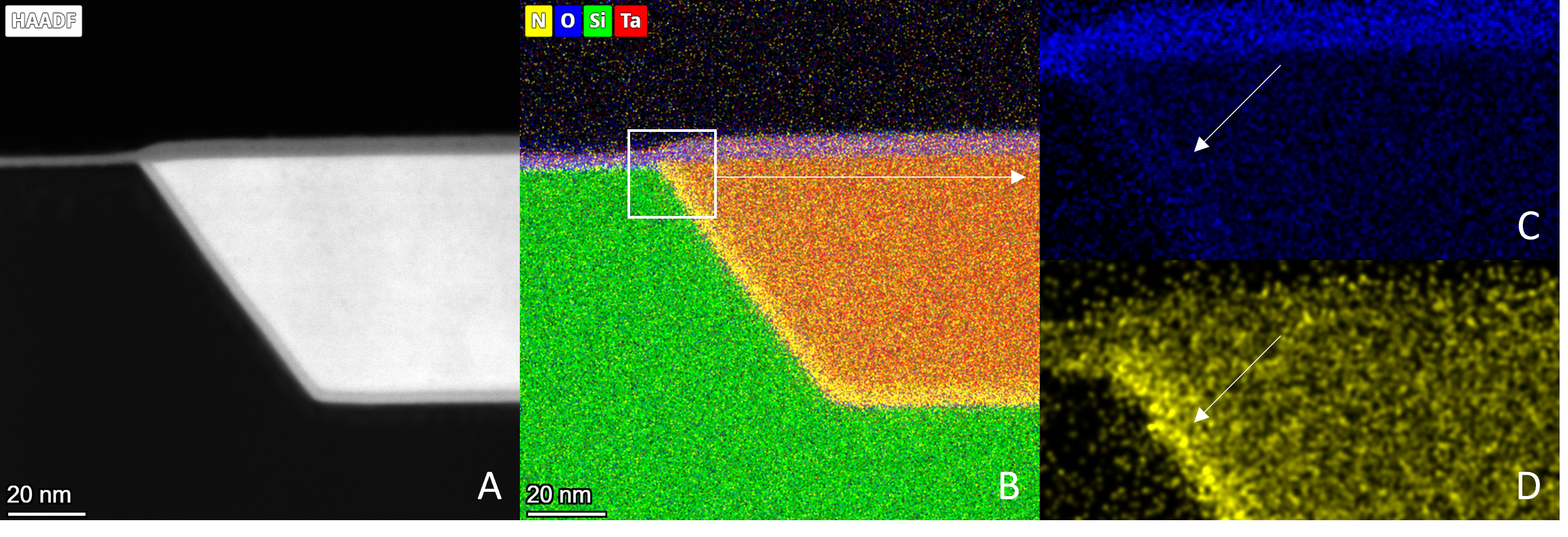}
\caption{\label{fig:Rep_B} {(A) STEM image of a Device B type showing clearly defined features relative to the Si substrate. (B) Composite elemental analysis of the device showing the N, O, Si, and Ta. Due to the overlap of the Si-K and Ta-M lines the Ta region is tinted from limitation on elemental identification. Enhanced image of the boxed area showing (C) the trapped oxygen and (D) N layer at the Si interface. The $\sim$3 nm of TaN exposed to atmosphere (O from TaON) is also visible. }}
\end{figure}

\begin{figure}[H]
\includegraphics[width=\linewidth]{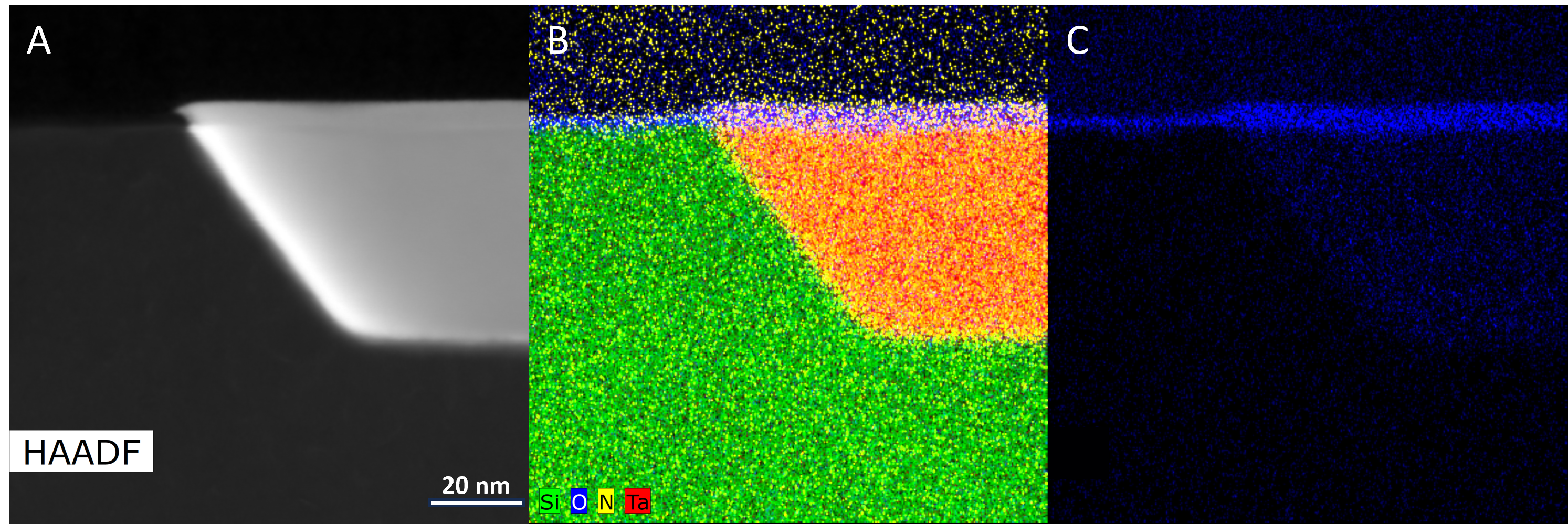}
\caption{\label{fig:Rep_C} {(A) STEM image of a Device C type showing clearly defined features relative to the Si substrate. (B) Composite elemental analysis of the device showing the N, O, Si, and Ta. Due to the overlap of the Si-K and Ta-M lines the Ta region is tinted from limitation on elemental identification. (C) Oxygen elemental analysis indicates no oxide at the Si/Ta interfaces. The 1~nm of TaN is not visible.}}
\end{figure}

\begin{figure}[H]
\includegraphics[width=\linewidth]{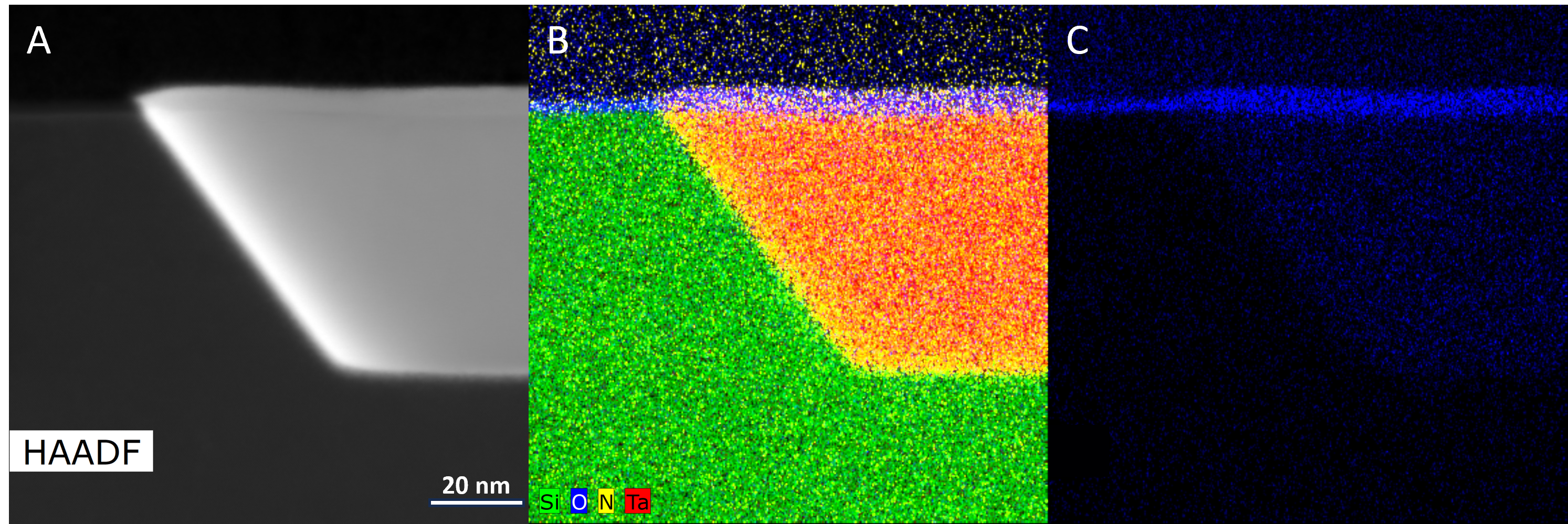}
\caption{\label{fig:Rep_D} {(A) STEM image of a Device D type showing clearly defined features relative to the Si substrate. (B) Composite elemental analysis of the device showing the N, O, Si, and Ta. Due to the overlap of the Si-K and Ta-M lines the Ta region is tinted from limitation on elemental identification. (C) Oxygen elemental analysis indicates no oxide at the Si/Ta interfaces. The 6~nm of TaN is clearly visible.}}
\end{figure}

\begin{figure}[H]
\includegraphics[width=\linewidth]{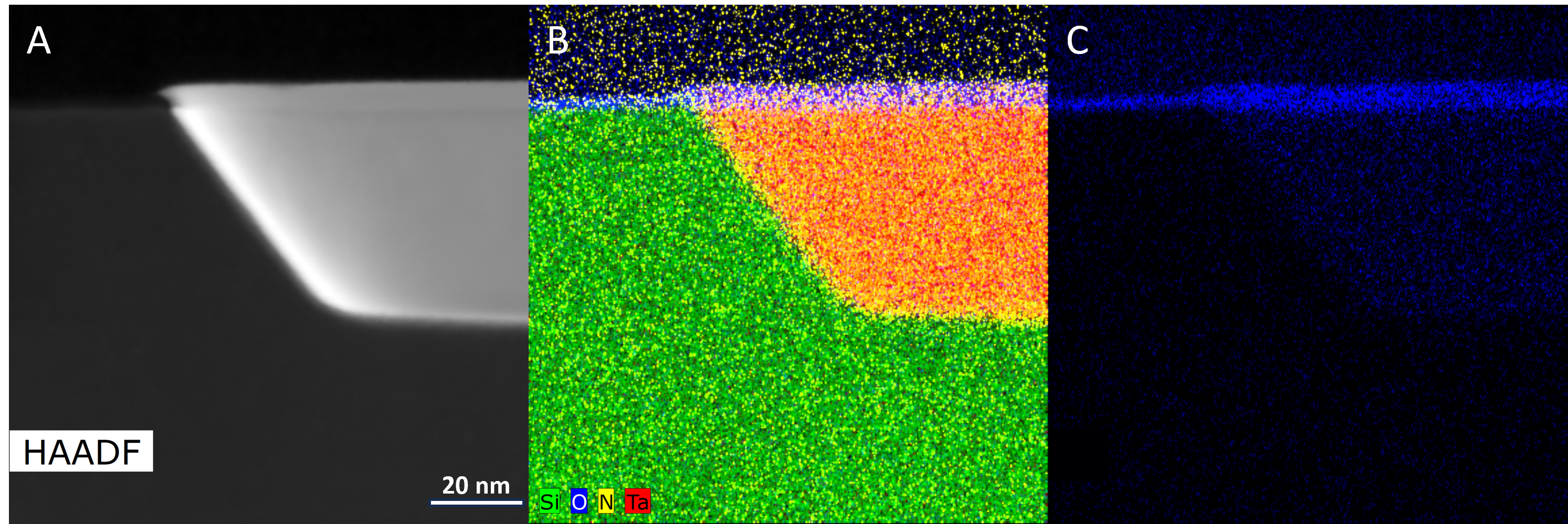}
\caption{\label{fig:Rep_E} {(A) STEM image of a Device E type showing clearly defined features relative to the Si substrate. (B) Composite elemental analysis of the device showing the N, O, Si, and Ta. Due to the overlap of the Si-K and Ta-M lines the Ta region is tinted from limitation on elemental identification. (C) Oxygen elemental analysis indicates no oxide at the Si/Ta interfaces. The 2~nm of PVD is present but appears diffused.}}
\end{figure}

An scanning transmission electron microscope (STEM) coupled with energy dispersive spectroscopy (EDS) is used to characterize the Ta, Si, N, and O elemental abundances in the resonator.  Representative STEM images and EDS maps from each set of devices are shown in Figures \ref{fig:Rep_A} - \ref{fig:Rep_E}. In each sample, the interfaces are clearly defined with the Si(100) plane visible in the high angle annular dark field (HAADF) image. The damascene thickness of the Ta is nominally the same at all points on the chip ($\sim$60~nm for Devices A, B and D, and $\sim$48 nm for Devices C and E); however, we note that devices C, D, adn E are thinner near the transmission line ($\sim$20~nm, not shown), which we believe affects the effective T$_C$ of these devices but not the quality factors.

For the Buried Oxide devices, both Devices A and B the N abundance overlaps with the first 10~nm of Ta near the Si interface, as expected, and is visible as a discoloration in the HAADF scan. Oxygen is mainly in the Si below the nitrogen layer with a minor band overlapping with the 6~nm nitrogen layer consistent with breaking vacuum and two 3~nm ALD TaN depositions. there is an additional O$_2$ layer at the Si/Ta interface, which we attribute to native oxide regrowth during a several hour fabrication delay. Additionally, the 2~nm TaN capping layer in Device B is not fully visible on the surface of the Si substrate likely obscured by limitation of  the elemental identification of the system. The CMP TaO$_x$ of both are $\leq$5~nm.

For the Pristine devices, Device C and D (ALD TaN deposition) and Device E (PVD TaN deposition) the N$_2$ is at the Ta/Si interface. The CMP TaO$_x$ is $\leq$5~nm with no additional oxides elsewhere. 

\subsection{Cryogenic Resonator Characterization}
The 5$\times$5~mm$^2$ damascene Ta resonator devices were individually adhered to an Au plated Cu puck using GE varnish. The puck has a recessed hole below the device back side to mitigate the coupling between the transmission line and the ground plane. The puck, along with another flat Au plated Cu disk, sandwiched a shielded PCB providing an RF enclosure for the device. The device was electrically connected to the PCB traces using an alloy wire Al (99\%)/Si (1\%) 25 $\mu$m with 2-3 wires bonding each input/output of the transmission feed-line and $\approx$30 wires between the PCB ground and the chip ground plane. The packaged devices were mounted onto an Au plated Cu cold finger attached to the mixing chamber plate of a BlueFors LD400 dilution refrigerator (DR) with a 10~mK base temperature. A Cryoperm magnetic shield enclosed the cold finger and doubled as an IR shield to the Still IR shield. 

\begin{figure}[H]
\includegraphics[width=\linewidth]{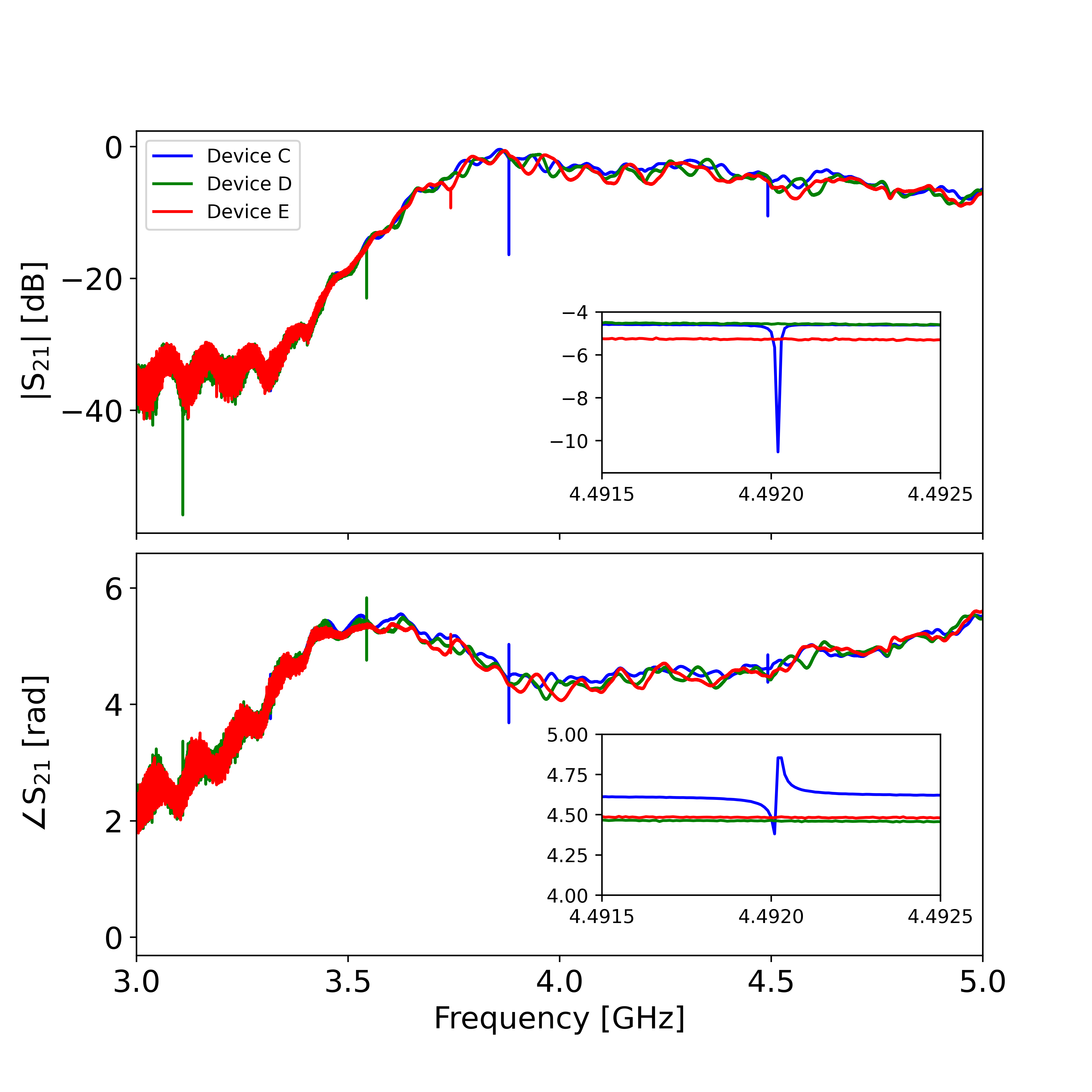}
\caption{\label{fig:Full_Scan} Full scan of (top) magnitude and (bottom) phase of \s. The drop in gain between 3 and 4~GHz is from the cryogenic circulator. Insets show enhanced image around a single resonator for Device C.}
\end{figure}

The input [output] of the PCB is connected to RF IR filters (from Quantum Microwave) which connect to other components on the MXC plate via hand formed tin plated Cu RF coax cables. On the input side, we use stainless steel housing attenuators at the 50~K (0~dB), 4~K (-20~dB), Still (-10~dB) plates and we use Cu housing attenuators at the CP (-10~dB) and MXC (-20~dB) plates to thermalize the semi-rigid sCuNi coax signal lines. On the output side, two LNF double circulators with cryo-terminators isolate the resonator from IR originating at warmer stage on the semi-rigid sCuNi coax signal lines. The return side signal is amplified at 4~K with a LNF HEMT amplifier by a nominally constant 39 dB in the 3 - 8 GHz bandwidth. We use flexible DELFT RF lines to connect the LNF HEMT to the 50~K stage returning to semi-rigid sCuNi coax to the output of the DR. At room temperature, two ZX60-123S amplifiers provide an additional total amplification of nominally 34~dB over the same bandwidth prior to readout by a VNA [Agilent E5071c]. An initial sweep of the S$_{21}$ response was taken from 3 to 5 GHz in 10~kHz steps with a 100~Hz IFBW for all devices to identify resonator tones at base temperature, Figure \ref{fig:Full_Scan}. The limitations of our readout chain is visible with a roll-off below 4~GHz due to the isolators used to mitigate IR and noise from the return side. 

\begin{figure}[H]
\includegraphics[width=\linewidth]{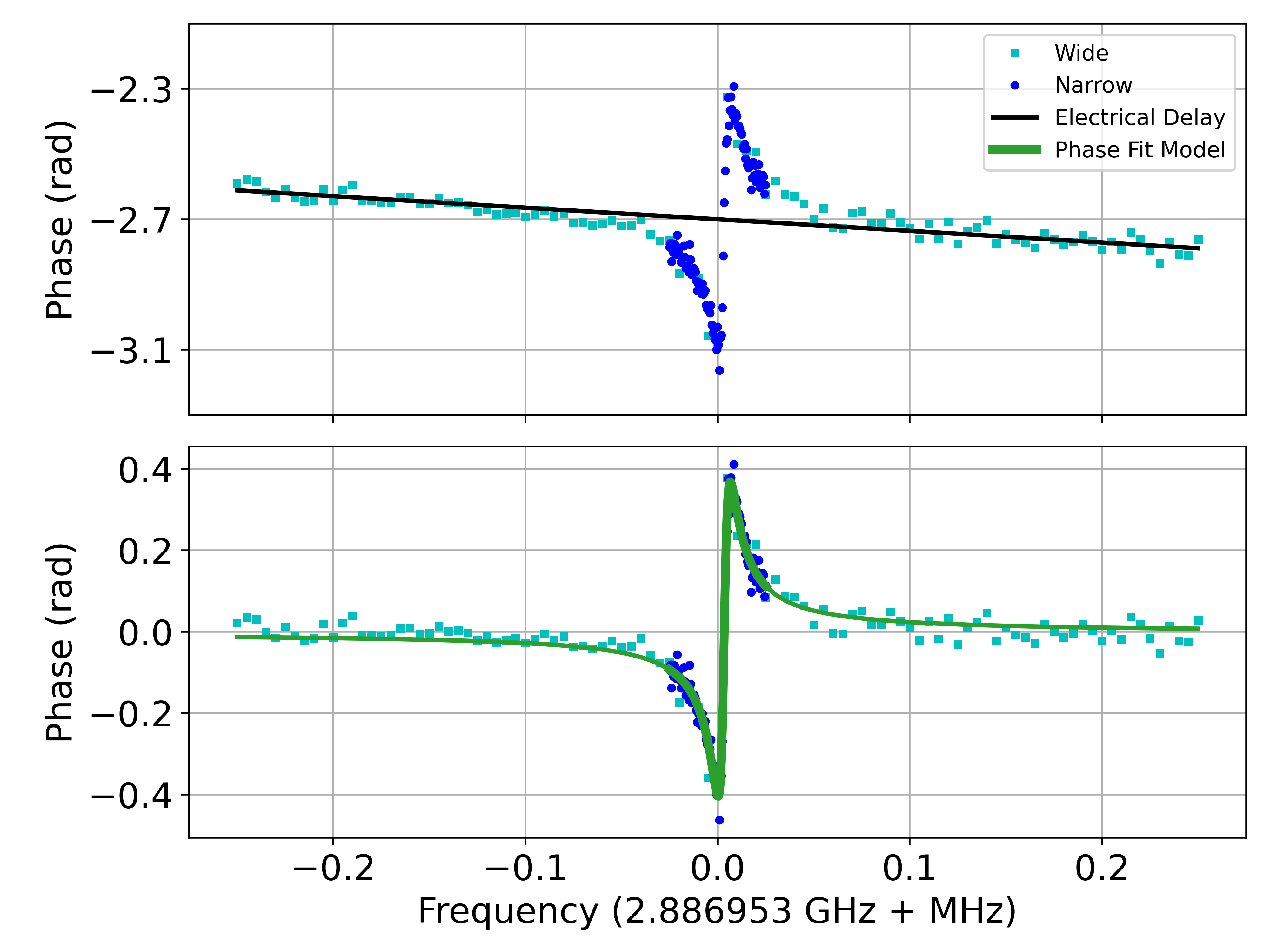}
\caption{\label{fig:Phase_fit} (Top) Raw data from VNA for both narrow and wide band scans. The electrical delay is fitted to the wide band data excluding the middle region ($\times$3 width of narrow band). (Bottom) The phase model is fitted to the background corrected data. The thick gold curve indicates the region where the fit is performed, while the narrow gold curve shows the extrapolation of the fit to the wide-band data, with good agreement observed.}
\end{figure}

A heater on the DR MXC plate was used to warm the DR from 25 to 600 mK in increments of 25 – 50~mK. Resonators were scanned at each step as a function of frequency at various powers (acquisition times vary from a few seconds to several minutes). A wide band scan ($\approx10 Q$) was performed to extract the electrical delay of the signal lines centered at each tone. The electrical delay was subtracted from the \s phase of a narrow band scan ($\approx2 Q$) centered on each tone, Figure \ref{fig:Phase_fit}. The resonator properties were extracted by fitting the background corrected phase of the \s data using the model from \cite{Khalil2012} modified for a constant offset,
\begin{equation}
    \phi(\omega) = \arg\left(1 - \frac{Q}{Q_C}\cdot\frac{1 + 2iQ\frac{\delta\omega}{\omega_0}}{1+2iQ\frac{\omega-\omega_0}{\omega_0}}\right) + \phi_0
\label{eqn:khalil_mag}
\end{equation}
where $Q$ is the total quality factor, $Q_C$ is the coupling quality factor, $\delta\omega$ is the asymmetry from the line impedance mismatch and $\phi_0$ is an offset. 

\section{Results}
\begin{table*}[t]
\centering
\begin{tabular}{|l|l|l|l|l|l|l|l|l|}
\hline
Position & Model & Device A      & Device B$_1$ & Device B$_2$ & Device B$_3$ & Device C  & Device D & Device E \\ \hline
1 (L)    & 4.2449& ---           & ---          & ---          & ---          & 2.8869552(3) & 3.1095532(9) & ---   \\ \hline
2 (C)    & 4.7440& 3.2927021(1)$^*$   & 3.36243(5)   & 3.3426635(8) & 3.292418(2)  & 3.3167549(4) & 3.5438650(1) & 3.1892495(7)\\ \hline
3 (L)    & 5.6803& 3.933598(2)   & 4.0359282(8) & 3.9886998(3) & 3.8579864(4) & 3.8800269(1) & ---          & 3.7426150(2)\\ \hline
4 (C)    & 6.6598& 4.4501541(4)  & 4.7164314(7) & 4.7868947(9) & ---          & 4.4920176(1) & ---          & ---  \\ \hline
\end{tabular}
\caption{Measured resonator tones for all devices at -60 dBm and 25 mK. *was taken at -50 dBm due to statistics limitations.}
\label{tab:Res_Table}
\end{table*}

A total of seven samples were tested consisting of Device A ($\times1$), B ($\times3$, separate locations on the wafer), C ($\times1$), D ($\times1$) and E ($\times1$) under nominally identical conditions. For Devices A and B the total attenuation at various stages was -60~dB. For Devices C - E the attenuation at various stages was reduced to a total of -30~dB with the installation of a cryogenic RF switch to measure the power delivered to the devices at base temperature. During measurements, an additional -30~dB of attenuation was attached at the room temperature input to the DR to keep the total attenuation constant. We recognize that the IR leakage through the lines was not identical between the two configurations. This introduces additional noise to the Devices C - E measurements but not impact the resonance tone or kinetic inductance.  

Under cryogenic testing 19 of the 28 possible resonators across the seven devices were observed. The most commonly unobserved resonance corresponded to the lower frequency inductively coupled resonator. The surfaces of devices were investigated using a helium ion microscope but no fabrication irregularities were observed.  

\begin{figure}[H]
\includegraphics[width=0.8\linewidth]{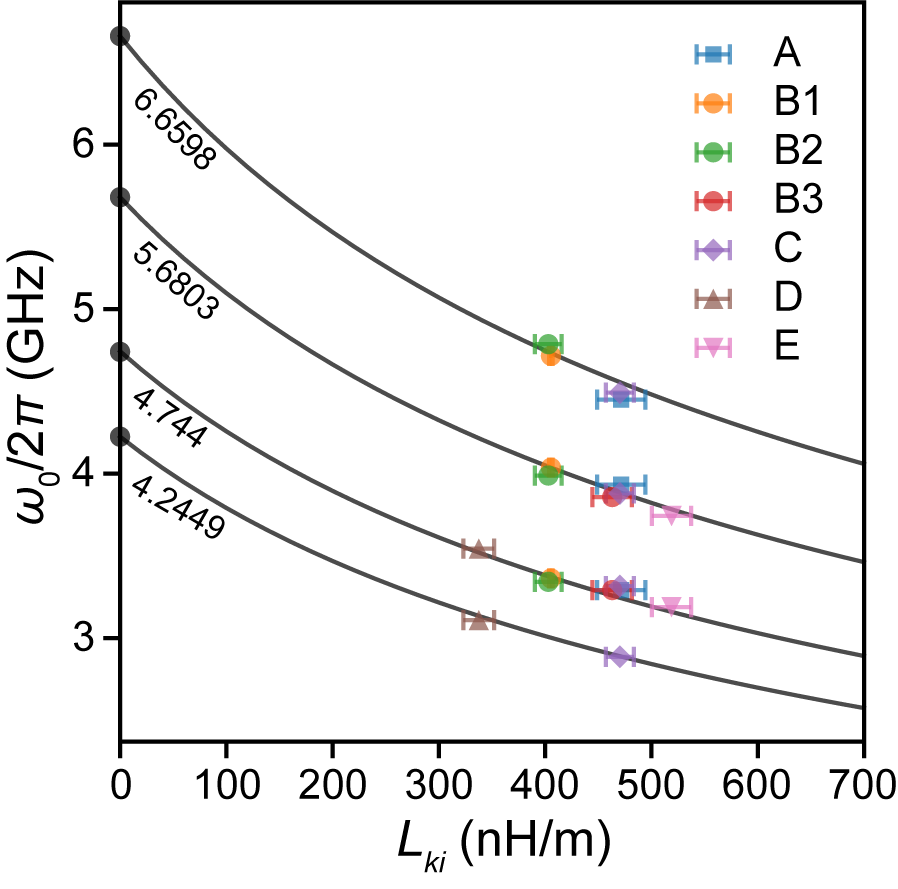}
\caption{\label{fig:KI_correction} \textbf{Simulated} resonance frequencies ($\omega_0/2\pi$) as a function of kinetic inductance $L_{ki}$. The circles at $L_{ki}=0$ denote the numerically simulated frequencies without kinetic inductance, while the solid lines show the corrected frequencies including $L_{ki}$. For each device set, $L_{ki}$ is extracted by matching the measured resonance frequencies to the simulated $\omega_0 - L_{ki}$ dependence.}   
\end{figure}

The resonance tones of our devices ($\omega_0^{e}$) were considerably lower than the theoretical values ($\omega_0^{t}$), Figure \ref{fig:KI_correction}, indicating a considerable contribution from the kinetic inductance $L_{ki}$ \cite{Li2024}, which follows the relation $\omega_0^{e}/\omega_0^{t}=\sqrt{L_m/(L_m+L_{ki})}$. The geometric inductance is calculated as 410 nH/m, following Gao’s approach \cite{Gao2008}. The kinetic inductance values were in a range that is not expected to influence the quality factor of our devices.

\begin{figure}[H]
\includegraphics[width=1.0\linewidth]{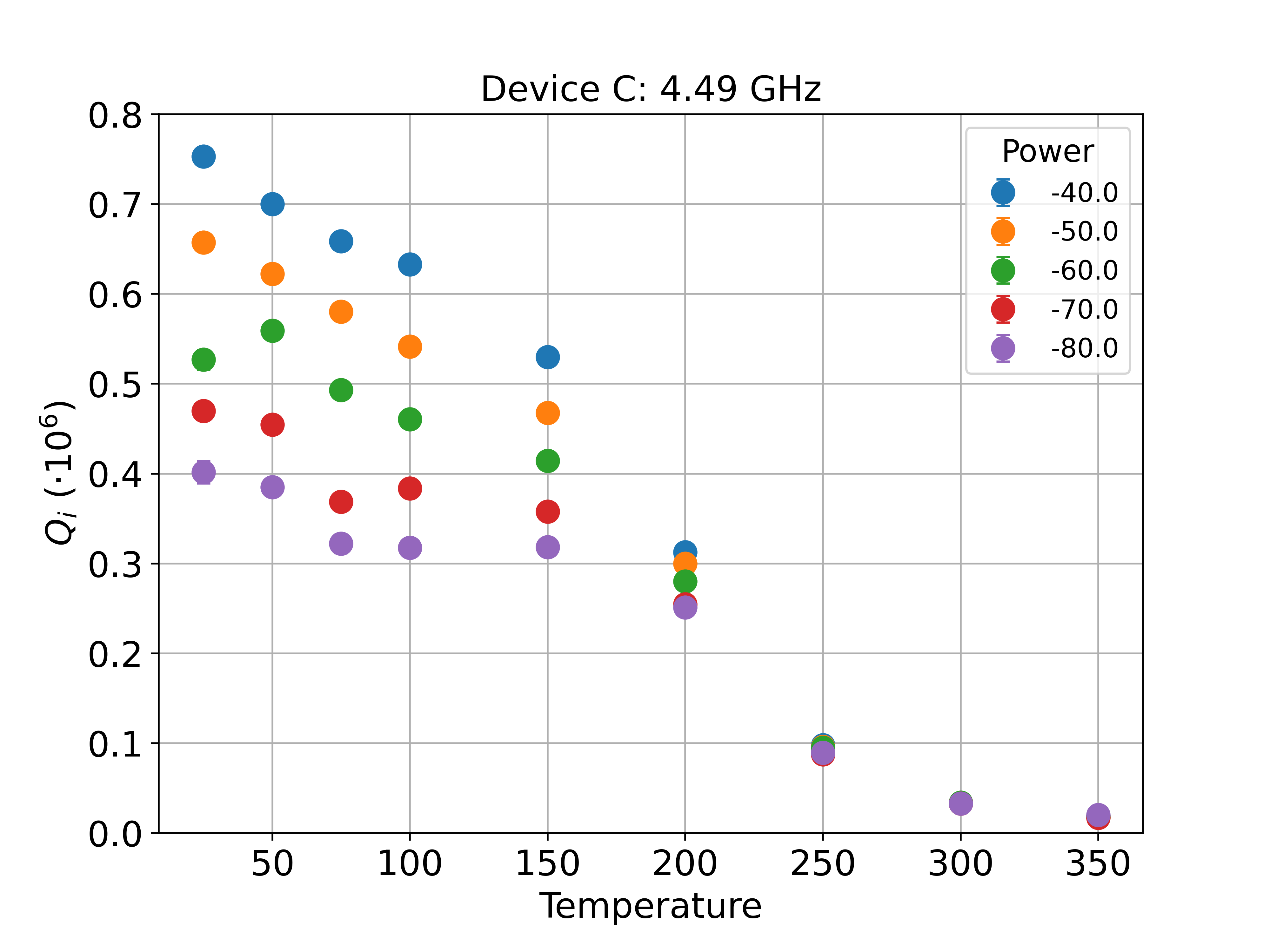}
\caption{\label{fig:Qi_T} Internal quality factor as a function of Temperature over five input powers. We see the effects of TLS at low input power and temperature and saturation for all powers at higher temperature.}
\end{figure}

\begin{figure}[H]
\includegraphics[width=1.0\linewidth]{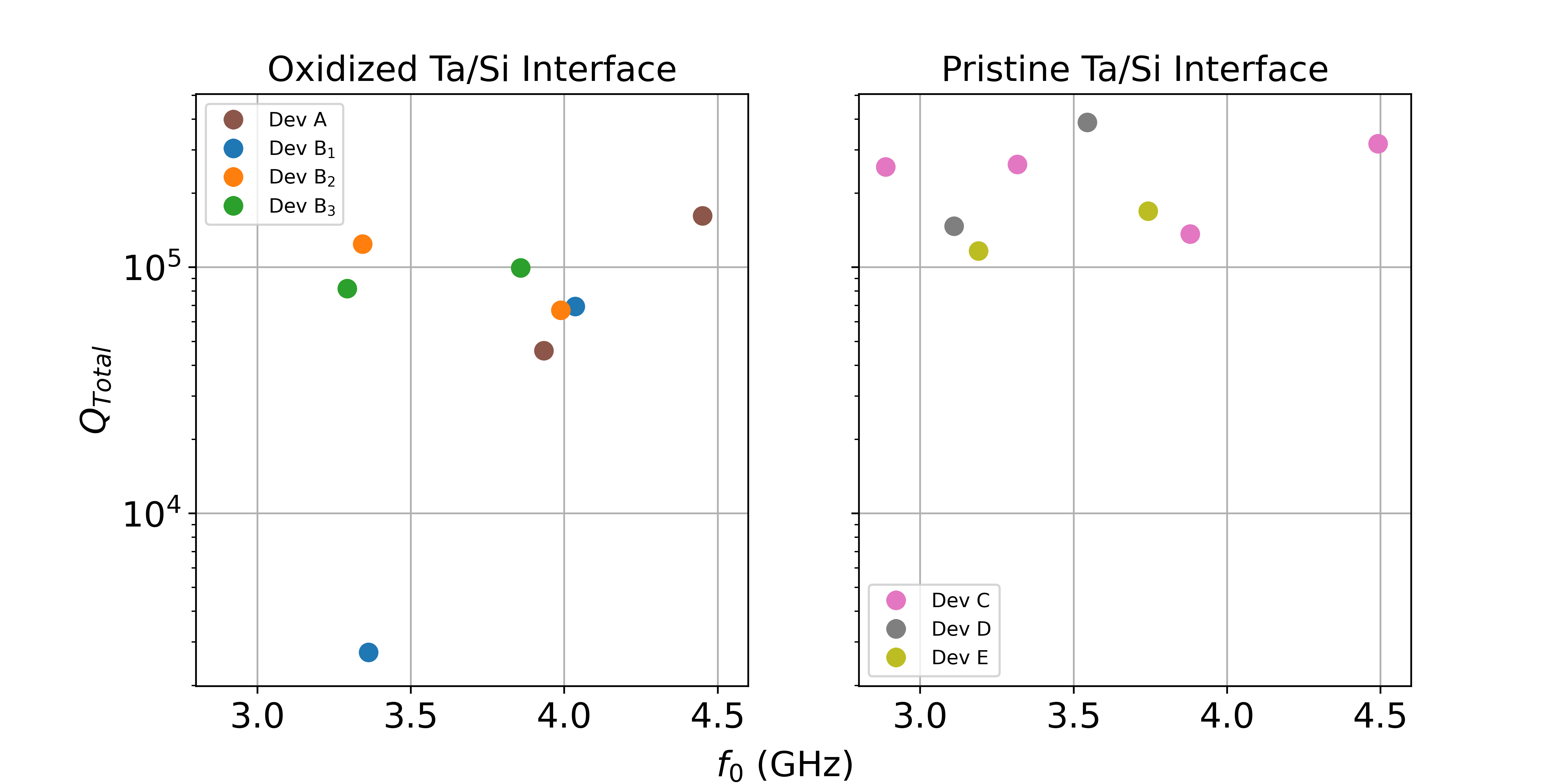}
\caption{\label{fig:Q-60} Quality factor for the seven devices under test at 25~mK and -60 dBm input power. Devices A and B have the additional oxide layer buried at the Si/Ta interface that contribute to degradations of their performance.}
\end{figure}

For the best performing resonator (Device C, 4.49 GHz) above 4~GHz to account for readout limitation, we measure a constant $Q_C$ of $(8.178 \pm 0.016) \times10^5$ over all considered temperature (25 to 350 mK) and power ranges (-40 to -80 dBm). The internal quality factor, Q$_I$, has the same three regimes associated with TLS coupling/saturation at 1) high temperatures and decoupled from power, 2) moderate temperatures and highly power dependent, and 3) low temperature and exponential dependence with power, as observed by others \cite{Crowley2023}, Figure \ref{fig:Qi_T}. Comparing the oxidized Ta/Si interface (Devices A and B) with the pristine Ta/Si (interface Devices C, D, and E) at -60~dBm, Figure \ref{fig:Q-60}, we observe a statistical improvement of $\approx$~$\times$2. However, a single resonator (Device A, 3.29~GHz) at -50 dBm performed comparably with pristine resonators; this suggests that there are unmitigated loss mechanisms that were affecting resonator performance beyond the Si/Ta interface. Additionally, we observed a smaller effective T$_C$ for devices C - E as defined by the early onset of TLS saturation with temperature. 

\section{Conclusions}
We demonstrated the application of the damascene process to fabricate resonators. The damascene process was used to convert the metal/air interface of the device sidewall into a metal/substrate interface. COMSOL MultiPhysics\textregistered~ simulations were consistent with experimental measurements after accounting for a higher kinetic inductance in Ta films. The devices were characterized across several temperatures and powers, with the effects of TLS observable at lower temperatures. Device performance between Pristine and Buried Oxide showed a modest improvement; however, thre is still significant variance in device performance in terms of the internal quality factor.

\section*{Acknowledgments}
We thank Patrick Truitt (Seeqc, Inc.) for the resonator design. This work was supported by the U.S. Department of Energy, Office of Science, National Quantum Information Science Research Centers, Co-design Center for Quantum Advantage under contract number DE-SC0012704 and PNNL FWP 76274. The numerical simulation used Theory \& Computation facility of the Center for Functional Nanomaterials, which is a U.S. Department of Energy Office of Science User Facility, at Brookhaven National Laboratory under Contract No. DE-SC0012704. 

\clearpage
\bibliography{references}

\end{document}